\documentclass[prx,aps,reprint,showpacs,amsfonts,amsmath,floatfix,superscriptaddress,footinbib]{revtex4-2}

\usepackage[svgnames]{xcolor}
\usepackage[acronym]{glossaries} 

\usepackage[colorlinks=true,        
            allcolors = black,  
            citecolor=DarkBlue, 
            linkcolor=DarkBlue,
            urlcolor=DarkBlue
        ]{hyperref} 

\usepackage{amsmath, amsfonts, amssymb, amsthm, bbm}
\usepackage{dsfont}
\usepackage{bm} 
\usepackage{mathtools}
\usepackage{physics}
\usepackage{graphicx}
\graphicspath{{pics/}}
\usepackage{dcolumn}
\usepackage{array}   
\usepackage{tikz}
\usepackage{comment}
\usetikzlibrary{quantikz}
\usepackage{dcolumn}
\usepackage{floatrow}
\usepackage[caption=false]{subfig}
\usepackage{multirow}

\newcolumntype{L}{>{$}l<{$}} 

\newcommand{\numqubits}{M}
\newcommand{\maxkNN}{K}
\newcommand{\knn}{$k$-NN~}
\newcommand{\numaminoacids}{N}
\newcommand{\RealAmpl}{\texttt{RealAmplitudes}~}

\newcommand{\groundstateenergy}{E_{\text{gs}}}

\newacronym{vqa}{VQA}{variational quantum algorithm}
\newacronym{vqe}{VQE}{variational quantum eigensolver}
\newacronym{qaoa}{QAOA}{quantum approximate optimization algorithm}
\newacronym{hva}{HVA}{Hamiltonian variational ansatz}
\newacronym{hea}{HEA}{hardware-efficient ansatz}
\newacronym{qaa}{QAA}{quantum adiabatic algorithm}
\newacronym{nisq}{NISQ}{noisy intermediate-scale quantum}
\newacronym{qubo}{QUBO}{quadratic unconstrained binary optimization}
\newacronym{psp}{PSP}{protein structure prediction}
\newacronym{fcc}{FCC}{face-centered cubic}
\newacronym{bcc}{BCC}{body-centered cubic}
\newacronym{hp}{HP}{hydrophobic-polar}
\newacronym{mj}{MJ}{Miyazawa-Jernigan}
\newacronym{mps}{MPS}{matrix product state}
\newacronym{hpc}{HPC}{high-performance computing}

\newacronym{average_error}{ARE}{average relative error}
\newacronym{best_case_error}{BCRE}{best-case relative error}

\begin{document}

\preprint{APS/123-QED}

\title{Efficient Quantum Protein Structure Prediction with Problem-Agnostic Ansatzes}

\author{Hanna Linn}
\email{hannlinn@chalmers.se}
\thanks{These authors contributed equally.}
\affiliation{Department of Microtechnology and Nanoscience (MC2), Chalmers University of Technology, G\"{o}teborg, 41296, Sweden}
\author{Rui-Hao Li}
\email{lir9@ccf.org}
\thanks{These authors contributed equally.}
\author{Alexander Holden}
\affiliation{Center for Computational Life Sciences, Cleveland Clinic, Cleveland, OH 44195, USA}
\author{Abdullah Ash Saki}
\affiliation{IBM Quantum, IBM Thomas J Watson Research Center, Yorktown Heights, NY 10598, USA}
\author{Frank DiFilippo}
\affiliation{Department of Nuclear Medicine, Cleveland Clinic, Cleveland, OH 44195, USA}
\author{Tomas Radivoyevitch}
\affiliation{Department of Quantitative Health Sciences, Cleveland Clinic, Cleveland, OH 44195, USA}
\author{Daniel Blankenberg}
\affiliation{Center for Computational Life Sciences, Cleveland Clinic, Cleveland, OH 44195, USA}
\author{Laura García-Álvarez}
\affiliation{Department of Microtechnology and Nanoscience (MC2), Chalmers University of Technology, G\"{o}teborg, 41296, Sweden}
\author{G\"{o}ran Johansson}
\affiliation{Department of Microtechnology and Nanoscience (MC2), Chalmers University of Technology, G\"{o}teborg, 41296, Sweden}

\begin{abstract}
Accurately predicting protein structures from amino acid sequences remains a fundamental challenge in computational biology, with profound implications for understanding biological functions and enabling structure-based drug discovery. 
Quantum computing approaches based on coarse-grained lattice models combined with variational algorithms have been proposed as an initial step towards predicting protein structures using quantum computers.
In this work, we introduce a more efficient quantum protein structure prediction workflow that bypasses the need for explicit Hamiltonian construction by employing a problem-agnostic ansatz.
The ansatz is trained to minimize an energy-based cost function that can be efficiently computed on classical computers, eliminating the need for ancillary qubits and reducing circuit depth compared to previous Hamiltonian-based methods.
This enables a more scalable approach for larger proteins and facilitates the inclusion of higher-order interactions, previously hard to achieve in quantum approaches. 
We validate our method by benchmarking a hardware-efficient ansatz on a large set of proteins with up to 26 amino acids, modeled on the tetrahedral, body-centered cubic, and face-centered cubic lattices, incorporating up to second-nearest-neighbor interactions.
We assess the performance on both a noise-free simulator and the \texttt{ibm\_kingston} quantum computer using a set of distinct metrics to probe different aspects of the prediction quality.
These experiments push the boundaries of quantum methods for protein structure prediction, targeting sequences that are longer than those typically addressed in prior studies.
Overall, the results highlight the scalability and versatility of our approach, while also identifying key areas for improvement to inform future algorithm development and hardware advancements.
\end{abstract}

\maketitle


\section{Introduction \label{sec:intro}}

The \gls{psp} problem, i.e., to accurately predict the three-dimensional structures of proteins from their amino acid sequences, has long been a central challenge in computational biology, owing to the intricate relationship between structure and function in biomolecular systems.
Protein structures dictate molecular function, interaction networks, and cellular pathways, making them critical for understanding biological processes and enabling structure-based drug discovery~\cite{goodsell2000structural, cavasotto2007ligand}.
The primary goal is to infer, from a linear amino acid sequence, the folded three-dimensional conformation that the protein adopts in its native biological environment.
In the template-based modeling regime, this challenge was partially addressed by the development of AlphaFold2~\cite{jumper_highly_2021}, whose success was recognized with the 2024 Nobel Prize in Chemistry~\cite{NobelPrizeChemistry}.
However, many proteins, such as those with novel folds or intrinsically disordered regions, lack homologous structures in existing databases, necessitating \textit{de novo} prediction methods that do not rely on structural templates.
In contrast, physics-based approaches aim to predict protein structures by modeling the underlying biophysical forces that drive folding, which may provide insights into folding mechanisms and dynamics that are not captured by template-based methods.
While fully atomistic molecular dynamics (MD) simulations can, in principle, model protein folding pathways, they are constrained by the exponential scaling of conformational space and the long timescales associated with biologically relevant folding events~\cite{shaw2010atomic, lindorff2011fast}.
Other popular classical methods, such as simulated annealing~\cite{kirkpatrick1983optimization, simons1997assembly} and conformational space annealing~\cite{lee1998conformational}, also face challenges in efficiently exploring the vast conformational landscape of proteins.

Quantum computing presents a fundamentally new approach for addressing the computational complexity inherent in \gls{psp}, leveraging superpositions and entanglement to explore the vast conformational space of proteins. 
However, translating this potential advantage into practical gains remains a significant challenge due to the current state of quantum hardware, with limited qubit counts, short coherence times, and low gate fidelities.
Given these constraints, quantum optimization algorithms are typically implemented as hybrid heuristics or \glspl{vqa} that offload parts of computations to classical processors. 
Approaches such as quantum annealing~\cite{das2008colloquium} and \gls{qaoa}~\cite{farhi_quantum_2014} have been investigated for potential advantages over classical methods in many optimization problems, in terms of speed and solution quality~\cite{blekos2024review, abbas2024challenges}.
Empirical evaluation of these quantum heuristics in current devices is essential to assess their utility, identify bottlenecks, and guide algorithm design and integration with high-performance classical computers. 
In the context of \gls{psp} applications, coarse-grained models that reduce degrees of freedom while retaining essential physical features have played a fundamental role in the development of quantum \gls{psp} for near-term hardware~\cite{perdomo_construction_2008, perdomo-ortiz_finding_2012, babbush2014construction, babej_coarse-grained_2018, fingerhuth_quantum_2018, robert_resource-efficient_2021, irback_folding_2022, boulebnane2023peptide, chandarana_digitized_2023, liQuantumAlgorithmProtein2025}.

\begin{figure*}[t]
\centering
{\includegraphics[width=0.95\textwidth]
{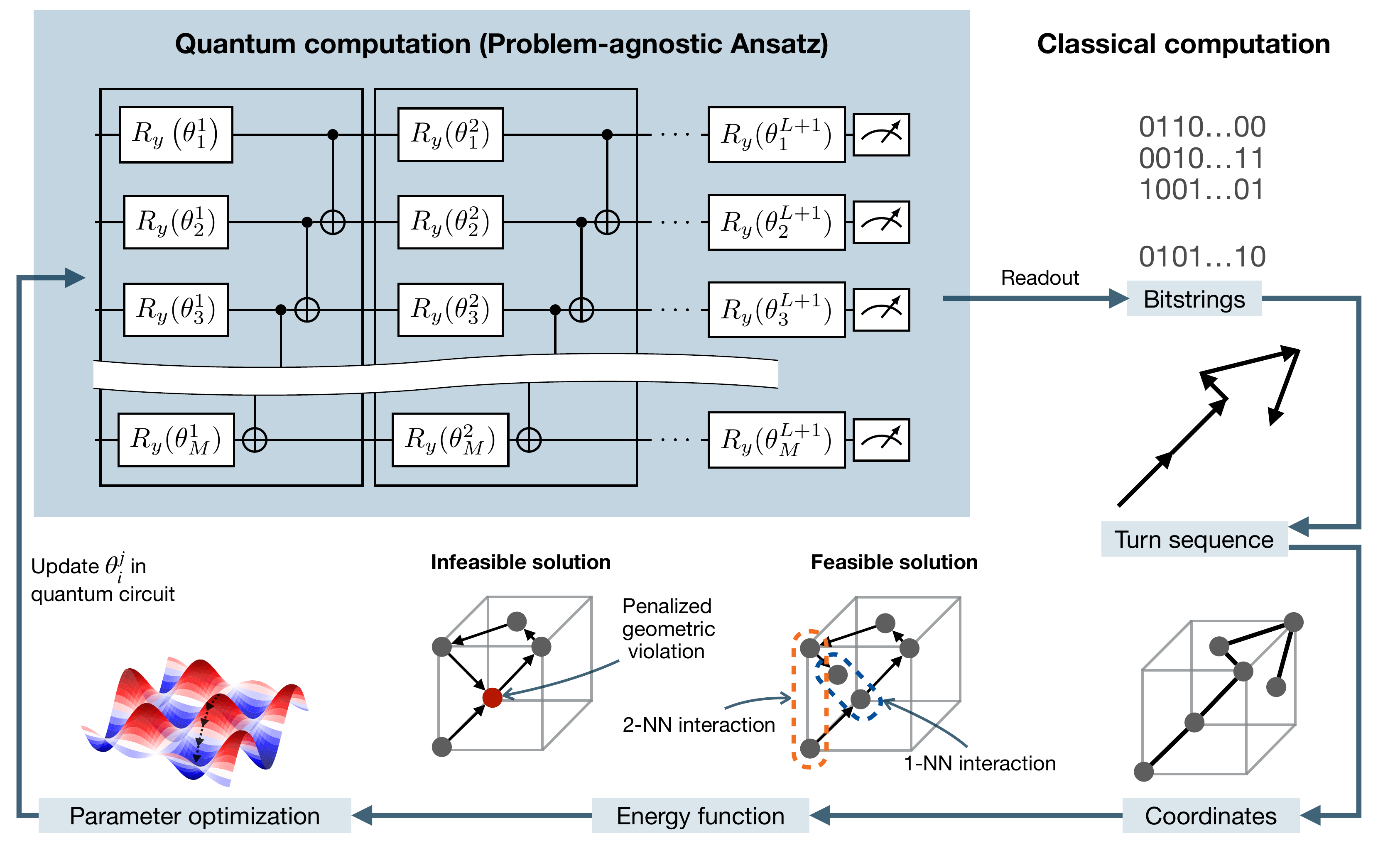}
}
\caption{\textbf{Overview of the Hamiltonian-free hybrid workflow for \gls{psp} using problem-agnostic ansatzes.} Bitstrings sampled from the variational quantum circuit are mapped to turn sequences, which are translated into spatial coordinates of candidate protein conformations. A classical energy function scores each configuration based on 1-NN (blue dashed line) and 2-NN (orange dashed line) interaction terms and geometric violations (red bead). The classical optimizer, e.g., COBYLA, iteratively updates the parameters $\theta_i^j$ of the quantum circuit with $\numqubits$ qubits and $L$ layers to minimize the energy, penalizing infeasible solutions and promoting feasible ones.
}
\label{fig:scheme}
\end{figure*}

The traditional workflow of quantum \gls{psp} typically involves the following steps.
First, a coarse-grained model is constructed to represent protein structures, where the amino acids are described by a reduced set of interaction sites, or beads, sitting on a discrete lattice.
These bead positions are then encoded into binary variables via either turn-based~\cite{perdomo-ortiz_finding_2012, babbush2014construction,babej_coarse-grained_2018, fingerhuth_quantum_2018,robert_resource-efficient_2021, boulebnane2023peptide} or coordinate-based encoding~\cite{perdomo_construction_2008,irback_folding_2022}, which are subsequently mapped to qubit operators.
Next, a Hamiltonian is constructed to represent the protein's energy landscape, typically including pairwise interactions and physical constraints, such as penalties associated with potential bead overlaps on the lattice.
Finally, a quantum algorithm is employed to find the ground state of this Hamiltonian, which corresponds to the lowest energy conformation of the protein.
The complexity of the Hamiltonian, and consequently the resource requirements, vary significantly depending on the lattice type and encoding scheme.
For example, ancillary qubits are required to implement the interaction terms in addition to the configuration qubits encoding the beads' positions. 
Typically, one ancilla qubit per amino acid pair is needed to flag whether they are first nearest neighbors, denoted 1-NN, thus making the total number of interaction qubits scale as \(O(\numaminoacids^2)\), where \(\numaminoacids\) is the number of amino acids in the protein. 
Extending beyond 1-NN interactions increases this overhead~\cite{robert_resource-efficient_2021}.
Moreover, for some lattices in turn-based encoding, such as the cubic lattice, the non-overlapping constraint requires more ancillary qubits if it is enforced through slack variables~\cite{babbush2014construction, babej_coarse-grained_2018}, leading to a qubit scaling of $O(\numaminoacids^2\log \numaminoacids)$.
In contrast, in lattices with simpler geometries, such as the tetrahedral lattice, such a constraint can be incorporated into the interaction Hamiltonian, thus avoiding the cost of slack variables and the associated qubit overhead~\cite{robert_resource-efficient_2021}.
A detailed resource analysis for coarse-grained quantum \gls{psp} with different encoding schemes and lattice geometries is provided in~\citet{linn2024resource}.

Beyond qubit count, the circuit depth of the quantum algorithm is also a critical factor that determines the feasibility of \gls{psp} on near-term quantum hardware.
In \glspl{vqa}, there are two main approaches to constructing the quantum circuit, or ansatz.
First, problem-specific ansatzes such as \gls{qaoa} embed the problem Hamiltonian into the quantum circuit by using a sequence of parameterized gates corresponding to its terms.
This approach can be efficient for certain problems, such as \gls{qubo} problems, where the resulting Hamiltonian is at most 2-local, with pairwise interactions between qubits.
However, for \gls{psp}, the Hamiltonian typically contains higher-order terms involving multi-qubit and long-range couplings, making it harder to directly embed the problem Hamiltonian into the circuit without incurring a large circuit depth.
Second, problem-agnostic ansatzes that are unaware of the structure of the specific problem, but are instead designed to be flexible and adaptable to various problems.
One such example is the \gls{hea}~\cite{kandala_hardware-efficient_2017}, which is typically a shallow circuit, consisting of alternating layers of single-qubit rotations and entangling gates that act on two or more qubits.
\gls{hea} can be seen as a universal approximator for quantum states, as the expressibility and the entangling capacity of the circuit can be adjusted by varying the number of layers and the connectivity of the qubits~\cite{sim2019expressibility}.
It has shown its ability to find the optimal protein conformations of small peptides constrained on a tetrahedral lattice~\cite{robert_resource-efficient_2021}.

In this work, we propose a resource-efficient method for performing quantum \gls{psp} that bypasses the Hamiltonian construction step, made possible by using a problem-agnostic ansatz, such as \gls{hea}.
As illustrated in Fig.~\ref{fig:scheme} and will be explained in detail in Sec.~\ref{sec:workflow}, the \gls{hea} is trained to minimize a cost function that is directly related to the energy of the protein conformation, which can be computed classically.
The main contributions of this work are twofold.
First, by removing the need to construct a Hamiltonian and encode it in the quantum circuit, we can avoid the previously mentioned qubit overhead and large circuit depths in conventional approaches.
In this approach, only configuration qubits are needed to encode the positions of the amino acids, with no additional ancilla qubits required for pairwise interactions or the non-overlapping constraint.
Moreover, due to the diagonal nature of the \gls{psp} Hamiltonian, all measurements can be performed in the computational basis, avoiding the measurement overhead typical in \glspl{hea} for quantum chemistry, where non-diagonal terms require basis rotations.
Altogether, this enables us to attempt \gls{psp} on larger proteins than previously possible.
Second, since the protein energy is processed classically, incorporating higher-order (\knn with $k>1$) pairwise interactions becomes straightforward and computationally efficient. 
In contrast, previous quantum computing implementations in the literature have mostly been limited to just 1-NN interactions. 
Methods that extend beyond 1-NN incur an exponential qubit overhead~\cite{robert_resource-efficient_2021}. 

In our numerical simulations, we demonstrate the effectiveness of this approach by predicting the structures of various proteins with up to 26 amino acids, modeled on three different lattices, the tetrahedral, \gls{fcc}, and \gls{bcc} lattices.
In contrast, earlier state-of-the-art quantum experiments on \gls{psp} have been limited to proteins with around 10 amino acids~\cite{chandarana_digitized_2023, robert_resource-efficient_2021, romero2025trappedionKipu}.
In addition, we showcase the results with higher-order interactions up to 2-NN. 
It should also be noted that the method can be easily extended to higher orders with little classical overhead.
After the variational quantum circuits are trained on a simulator, we run them on IBM quantum hardware with superconducting qubits to validate the performance of our approach in a noisy environment.
As a result, our strategy enables empirical testing of quantum algorithms on proteins much larger than previously attempted, using up to 46 qubits, testing current hardware capabilities beyond proof-of-principle demonstrations and supporting potential algorithm–hardware co-design.

The structure of this paper is as follows. 
Section~\ref{sec:lattice} introduces various lattice encodings used for protein structure representation, including tetrahedral, \gls{bcc}, and \gls{fcc} lattices. 
Section~\ref{sec:workflow} describes our Hamiltonian-free hybrid workflow, detailing the \gls{hea}, classical evaluation of the energy function, optimization strategies, and the performance metrics employed. 
Section~\ref{sec:results} presents our statevector and \gls{mps} simulation results, as well as results obtained on the \texttt{ibm\_kingston} quantum device, focusing on the energy distribution, the average relative error, and the best-case relative error. 
Finally, Section~\ref{sec:discussion} offers a discussion of the implications of our findings and outlines potential directions for future research.

\section{Lattice encodings}
\label{sec:lattice}

\begin{figure}[htpb]
    \centering
    \begin{minipage}[t]{0.48\columnwidth}
        \centering
        \sidesubfloat[]{\includegraphics[width=0.8\linewidth]{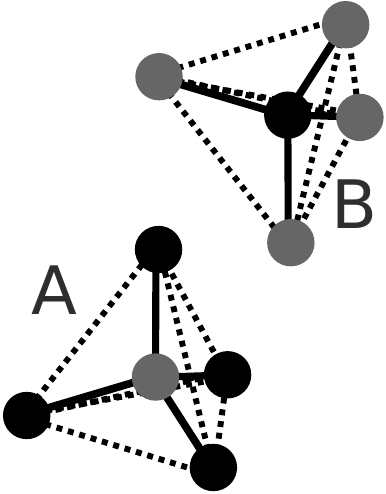}\label{fig:tetra}}
    \end{minipage}%
    \hfill
    \begin{minipage}[t]{0.48\columnwidth}
        \centering
        \sidesubfloat[]{\includegraphics[width=0.8\linewidth]{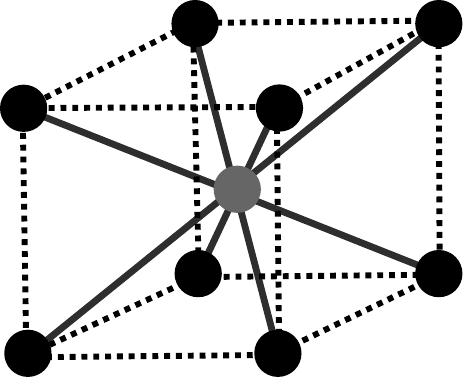}\label{fig:bcc}}
    \end{minipage}
    \vskip 2 em
    \begin{minipage}[t]{0.48\columnwidth}
        \centering
        \sidesubfloat[]{\includegraphics[width=0.8\linewidth]{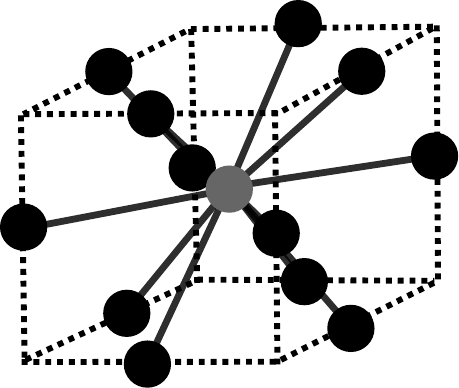}\label{fig:fcc}}
    \end{minipage}
\caption{\textbf{Unit cell representations of the three lattices used in this work: (a) tetrahedral, (b) \gls{bcc}, and (c) \gls{fcc}.} 
Each bead represents an amino acid in the coarse-grained protein model, and the solid lines indicate the possible turn directions between connected beads.
The tetrahedral lattice has a coordination number of 4, while the BCC and FCC lattices have coordination numbers of 8 and 12, respectively.
Note that the tetrahedral lattice can be divided into two sublattices, $A$ and $B$, as indicated by the different colors of the beads in (a). }
\label{fig:lattices}
\end{figure}

In coarse-grained quantum \gls{psp}, each amino acid on the protein chain is represented by a bead that occupies a discrete position on a lattice.
The lattice structure defines the spatial constraints and interaction topology of the protein.
Therefore, the choice of lattice geometry is crucial for accurately modeling protein structures.
In general, the more granular the lattice is, that is, the more degrees of freedom it has, the more accurately it is able to represent the realistic protein structures~\cite{godzik1993lattice}.
In the meantime, the increased granularity demands more quantum resources for encoding, as well as an increased computational cost for classical post-processing.
Finding the right balance between granularity and computational efficiency is therefore a key challenge in coarse-grained quantum \gls{psp}.
In this work, we test our approach on three different lattice geometries with increasing coordination numbers, i.e., the number of beads that are directly bonded to any given bead: the tetrahedral, \gls{bcc}, and \gls{fcc} lattices. 
See Fig.~\ref{fig:lattices} for an illustration of the unit cells of these lattices.
Below, we briefly describe the lattice geometries and their corresponding turn encodings to binary variables.

\subsection{Tetrahedral lattice}
\label{sec:tetra}

The tetrahedral lattice (Fig.~\ref{fig:tetra}) is a relatively simple lattice structure with a coordination number of four, that is, each bead can connect to four other beads.
In the context of turn-based encoding~\cite{babbush2014construction,babej_coarse-grained_2018}, this means that each bead can have four possible turn directions, which can be represented by two qubits in binary encoding.
On this lattice, only a single turn angle is permitted between any two consecutive turns, i.e., among three connected beads, corresponding to the tetrahedral bond angle of 109.47$^\circ$.
\citet{robert_resource-efficient_2021} first studied the quantum \gls{psp} problem on the tetrahedral lattice using an \gls{hea}. 
With binary encoding, the number of configuration qubits needed to encode the turns of a protein with \(\numaminoacids\) amino acids is $2 (\numaminoacids - 1) - 5$.
The extra reduction of five qubits comes from the rotational symmetry of the protein chain, which allows the first two turns and one of the qubits in the third turn to be fixed.
It is worth noting that the tetrahedral lattice is bipartite, meaning that the beads can be divided into two alternating sets, collectively labeled as the $A$ and $B$ sublattices, as seen in Fig.~\ref{fig:tetra}.
At the $A$ sublattice, each bead can only grow in the directions $t_i \in \{0, 1, 2, 3\}$, while at the $B$ sublattice, the beads can only take turns $t_i \in \{\bar 0, \bar 1, \bar 2, \bar 3\}$, where $\bar n$ is the opposite turn direction of $n$ (see Table \ref{tab:tetra_turns}).
The turn encoding of the tetrahedral lattice is given in Table~\ref{tab:tetra_turns}.

\begin{table}[h]
\caption{\label{tab:tetra_turns}\textbf{Turn encoding of the tetrahedral lattice.} The first four rows correspond to the $A$ sublattice, while the last four rows correspond to the $B$ sublattice. The turn vectors are written in Cartesian coordinates, where each vector represents the direction of the turn relative to the previous bead. Each turn direction is encoded using two qubits.}
\begin{ruledtabular}
\begin{tabular}{c c l c}
    & Turn label & Turn vector & Qubit encoding \\
    \hline
    \multirow{4}{*}{\hspace{1em}$A$} 
        & \rule{0pt}{1em}0 & $(1, 1, 1)$ & $00$ \\
        & 1 & $(-1, -1, 1)$ & $01$ \\
        & 2 & $(1, -1, -1)$ & $10$ \\
        & 3 & $(-1, 1, -1)$ & $11$ \\
    \hline
    \multirow{4}{*}{\hspace{1em}$B$} 
        & \rule{0pt}{1em}$\bar{0}$ & $(-1, -1, -1)$ & $00$ \\
        & $\bar{1}$ & $(1, 1, -1)$ & $01$ \\
        & $\bar{2}$ & $(1, -1, 1)$ & $10$ \\
        & $\bar{3}$ & $(-1, 1, 1)$ & $11$ \\
\end{tabular}
\end{ruledtabular}
\end{table}

\subsection{Body-centered cubic (BCC) lattice}
\label{sec:bcc}

The \gls{bcc} lattice (Fig.~\ref{fig:bcc}) is a more complex lattice structure with a coordination number of eight, which requires three qubits to encode each turn on the protein chain.
The higher coordination number of the \gls{bcc} lattice gives rise to more possible turn angles, $\{70.53^\circ, 109.47^\circ, 180^\circ\}$, making it more flexible than the tetrahedral lattice.
The turn encoding for this lattice is shown in Table~\ref{tab:bcc_turns}.
In this case, the number of configuration qubits is $3 (\numaminoacids - 1) - 4$, where the first turn and an additional qubit in the second turn can be fixed due to the rotational symmetry.
The use of \gls{bcc} lattice in the context of quantum \gls{psp} was first presented by~\citet{wong2021quantum}, where the authors employed Grover's search algorithm in conjunction with the \gls{hp} model to identify low-energy conformations.

\begin{table}[h]
\caption{\label{tab:bcc_turns}\textbf{Turn encoding of the \gls{bcc} lattice.}}
\begin{ruledtabular}
\begin{tabular}{c l c}
Turn label & Turn vector & Qubit encoding \\
\hline
       \rule{0pt}{1em}0 & \((1, 1, -1)\) & \(000\) \\
        1 & \((-1, 1, -1)\) & \(001\) \\
        2 & \((-1, -1, 1)\) & \(010\) \\
        3 & \((-1, 1, 1)\) & \(011\) \\
        4 & \((1, -1, -1)\) & \(100\) \\
        5 & \((1, -1, 1)\) & \(101\) \\
        6 & \((-1, -1, -1)\) & \(110\) \\
        7 & \((1, 1, 1)\) & \(111\) \\ 
\end{tabular}
\end{ruledtabular}
\end{table}

\subsection{Face-centered cubic (FCC) lattice}
\label{sec:fcc}

Finally, the \gls{fcc} lattice (Fig.~\ref{fig:fcc}) represents one of the most densely packed lattice structures, with a coordination number of 12.
Among the three lattices considered in this work, it has the highest degree of connectivity, resulting in the largest set of allowed turn angles: $\{60^\circ, 90^\circ, 120^\circ, 180^\circ\}$.
A clear advantage of the \gls{fcc} lattice is that it allows for a more accurate representation of secondary structures in proteins, particularly alpha helices.
In an alpha helix, the backbone of the protein forms a right-handed coil, with each amino acid residue contributing to a 100$^\circ$ turn along the helical axis, corresponding to 3.6 residues per helical turn~\cite{pauling1951structure}.
The tetrahedral lattice, with only one turn angle of 109.5$^\circ$, cannot reproduce the alpha-helix geometry, whereas the FCC lattice, with its richer set of discrete angles, can approximate the effective 100$^\circ$-per-residue twist through suitable sequences of turns.
For example, a sequence of (90$^\circ$, 120$^\circ$, 90$^\circ$) turns results in an average turn angle of 100$^\circ$.

When it comes to turn encoding, the \gls{fcc} lattice requires four qubits to represent each turn, leading to a total of $4 (\numaminoacids - 1) - 6$ configuration qubits.
It is worth pointing out that four qubits give a total of 16 possible states (bitstrings), but only 12 of them are used to represent the turn directions in the \gls{fcc} lattice.
The remaining four bitstrings do not correspond to any physical turn direction and therefore, a configuration containing any of these bitstrings needs to be penalized in the energy function (see more details in Sec.~\ref{sec:energy_function}).
Recent work by \citet{liQuantumAlgorithmProtein2025} studied the encoding of the \gls{fcc} lattice and the corresponding Hamiltonian construction for quantum \gls{psp}.
We adopt the same turn encoding scheme, which is shown in Table~\ref{tab:fcc_turns}.

\begin{table}[h]
\caption{\label{tab:fcc_turns}\textbf{Turn encoding of the \gls{fcc} lattice.}}
\begin{ruledtabular}
\begin{tabular}{c l c}
Turn label & Turn vector & Qubit encoding \\
    \hline
        \rule{0pt}{1em}0 & \((1, 1, 0)\) & \(0000\) \\
        1 & \((-1, -1, 0)\) & \(0011\) \\
        2 & \((-1, 1, 0)\) & \(1100\) \\
        3 & \((1, -1, 0)\) & \(1111\) \\
        4 & \((0, 1, 1)\) & \(1001\) \\
        5 & \((0, -1, -1)\) & \(0101\) \\
        6 & \((0, 1, -1)\) & \(1010\) \\
        7 & \((0, -1, 1)\) & \(0110\) \\ 
        8 & \((1, 0, 1)\) & \(1000\) \\
        9 & \((-1, 0, -1)\) & \(0100\) \\
        10 & \((1, 0, -1)\) & \(1011\) \\
        11 & \((-1, 0, 1)\) & \(0111\) \\
\end{tabular}
\end{ruledtabular}
\end{table}

\section{Hamiltonian-free hybrid workflow \label{sec:workflow}}

In this section, we present our Hamiltonian-free hybrid workflow for quantum \gls{psp} that leverages classical post-processing to compute the energy of protein conformations, which is illustrated in Fig.~\ref{fig:scheme}.
As mentioned in Sec.~\ref{sec:intro}, this approach provides two main advantages over traditional quantum \gls{psp} methods that rely on a Hamiltonian construction. 
First, it eliminates the need for ancillary qubits that may be required for encoding pairwise interactions and enforcing the non-overlapping constraint.
Second, it allows for a straightforward incorporation of higher-order interactions beyond first nearest neighbors, which can potentially lead to more accurate predictions of protein structures.
Moreover, the classical post-processing step can be accelerated by leveraging advanced \gls{hpc} infrastructures, together with efficient parallelization techniques, to handle the computationally intensive energy calculations for large protein instances.

One of the key components that makes this workflow possible is the use of a problem-agnostic ansatz.
Unlike problem-specific ansatzes such as \gls{qaoa}, which require explicit embedding of the protein Hamiltonian into the quantum circuit, problem-agnostic ansatzes such as the \gls{hea} do not assume any knowledge of the problem structure.
This flexibility allows us to decouple the encoding of the problem instance from the structure of the variational quantum circuit.
In our approach, the configuration of a protein---specifically, the spatial arrangement of its coarse-grained residues on a lattice---is directly encoded into a set of qubits representing the turns of the protein chain.
The \gls{hea} then acts on this register of configuration qubits, the resulting quantum state is measured, and each sampled bitstring is interpreted as a potential protein conformation.
Instead of encoding the energy landscape into a Hamiltonian and using the quantum circuit to estimate its expectation value, we evaluate the energy of the measured conformations classically, based on an energy function that accounts for pairwise interactions and constraints of the lattice model.
This enables us to define a cost function purely in terms of the classical energy, which is minimized by training the parameters of the \gls{hea} using a classical optimizer.
Since all energy calculations are offloaded to classical post-processing, the quantum circuit is not constrained by the depth or locality of the protein Hamiltonian, making the workflow more scalable and adaptable to different lattice types and interaction models.
In the following, we describe the details of the workflow, including the ansatz design, energy function construction, optimization details, and performance metrics used in the subsequent experiments.

\subsection{Hardware-efficient ansatz}
\label{sec:hea}
Problem-agnostic ansatzes are variational quantum circuits that are designed without incorporating problem-specific structures, making them adaptable to a wide range of problems.
One of the most widely used problem-agnostic ansatzes in the \gls{nisq} era is the \gls{hea}~\cite{kandala_hardware-efficient_2017}, which consists of alternating layers of single-qubit rotations and entangling gates arranged according to the hardware topology.
\glspl{hea} are particularly suitable for \gls{nisq} devices, as they minimize circuit depth and gate overhead by taking into account the hardware's connectivity, while still offering sufficient expressibility for many variational tasks~\cite{sim2019expressibility, cerezo_variational_2021}.
In this work, we use a variant of the \gls{hea} called \RealAmpl \cite{RealAmplitudesLatestVersion}, which is a single-layered circuit consisting of a layer of parameterized single-qubit rotations (such as $R_y$), followed by a layer of entangling gates (such as CNOT), and then another layer of parameterized single-qubit rotations before measurement.
An illustration of the circuit is shown in Fig.~\ref{fig:scheme}.
The single-qubit rotations are parameterized by angles $\theta_i$, which are optimized during the training process.
The entangling gates are arranged in a reverse linear fashion, meaning that the last qubit is entangled with the second-to-last qubit, which is then entangled with the third-to-last qubit, and so on.
In this case, the circuit depth as well as the CNOT-gate depth increase linearly with the number of qubits, and so does the number of trainable parameters in the circuit.

Therefore, in our Hamiltonian-free \gls{psp} workflow, we leverage this class of ansatzes to encode trial conformations of a protein into quantum states, sample bitstrings from the circuit, and compute their energies entirely with classical post-processing.
This decoupling of circuit architecture from energy modeling avoids the resource-intensive Hamiltonian construction step. 
It therefore enables a flexible incorporation of complex energy terms, including higher-order interactions and lattice-specific constraints, without requiring additional qubits.

\subsection{Energy function evaluation \label{sec:energy_function}} 

The main idea behind \textit{ab initio} \gls{psp} is to score all possible conformations of a protein chain based on their physical interactions and constraints, and then find the conformation with the lowest energy.
In this subsection, we introduce the core of our Hamiltonian-free quantum \gls{psp} workflow, which is the energy function that scores the sampled conformations based on their physical properties.
The energy function accounts for the pairwise interactions between beads, as well as the constraints imposed by the lattice structure.
It is defined in a similar way to a lattice Hamiltonian with turn-based encoding, such as the one used in~\citet{babbush2014construction} and~\citet{babej_coarse-grained_2018}.
Specifically, the energy function can be written in the form:
\begin{equation} \label{Eq:energy_func}
    E(\vb q) = E_{\text{olap}}(\vb q) + E_{\text{int}}(\vb q) + E_{\text{redun}}(\vb q),
\end{equation}
where $\vb q$ represents the qubits needed to encode the protein conformation, i.e., the configuration qubits.
The first term, $E_{\text{olap}}$, is the penalty for any overlapping beads, which ensures that no two beads occupy the same lattice site.
The second term, $E_{\text{int}}$, is the interaction energy term, which captures the pairwise interactions between beads based on their distances in real space.
Finally, $E_{\text{redun}}$ is an optional term that penalizes redundant turns, i.e., turns that do not correspond to any physical turn direction on a lattice.
Among the three lattices considered in this work, only the \gls{fcc} lattice requires this term, as its turn encoding generates more qubit states than the physical turn directions.

The redundancy term is the simplest one, as it only requires checking whether the sampled bitstring contains any of the redundant states, which are not physically allowed on the lattice.
With the \gls{fcc} encoding shown in Table~\ref{tab:fcc_turns}, the redundant states are $S_{\text{redun}} = \{0001, 0010, 1101, 1110\}$.
Then, for each of these states in the bitstring, we add a large positive penalty, $\lambda_{\text{redun}}$, to the energy function:
\begin{equation}
    E_{\text{redun}}(\vb q) = \lambda_{\text{redun}} \sum_{t=1}^{\numaminoacids - 1} f(\vb q^{(t)}),
\end{equation}
where the indicator function $f(\vb q^{(t)}) = 1$ if the $t$-th turn in the bitstring is one of the redundant states, $\vb q^{(t)} \in S_{\text{redun}}$, and $f(\vb q^{(t)}) = 0$ otherwise.

The other two terms hinge on computing the pairwise distances between beads in the protein chain.
With the turn-based encoding, the position of each bead can be easily computed from the position of the previous bead and the turn direction.
For example, if the position of the $i$-th bead is given by $\vb r_i$ in Cartesian coordinates, then the position of the $(i+1)$-th bead can be computed as $\vb r_{i+1} = \vb r_i + \vb t_i$, where $\vb t_i$ is the turn direction of the $i$-th bead, which can be obtained from the corresponding qubit state in the bitstring.
The set of turn directions for the three lattices is listed in the second column of Tables~\ref{tab:tetra_turns}, \ref{tab:bcc_turns}, and \ref{tab:fcc_turns}.
The distance between two beads can then be computed as the Euclidean distance between them:
\begin{equation}
    \begin{split}
    d_{ij} & = \left\| \vb r_i - \vb r_j \right\| \\
     & = \sqrt{(x_i - x_j)^2 + (y_i - y_j)^2 + (z_i - z_j)^2},
    \end{split}
\end{equation}
where $(x_i, y_i, z_i)$ and $(x_j, y_j, z_j)$ are the Cartesian coordinates of the $i$-th and $j$-th beads, respectively.
Then, for a protein with $\numaminoacids$ amino acids, we compute the pairwise distances between all pairs of beads, resulting in a symmetric distance matrix $\vb D \in \mathbb{R}^{\numaminoacids \times \numaminoacids}$.
To establish the penalty term for overlaps, we check all entries in the upper triangle of the distance matrix $\vb D$---if any two beads overlap, i.e., $d_{ij} = 0$, we add a large positive penalty, $\lambda_{\text{olap}}$, to the energy function.
Therefore, the full overlap energy term is written as
\begin{equation}
    E_{\text{olap}}(\vb q) = \lambda_{\text{olap}} \sum_{i=1}^{\numaminoacids - 1} \sum_{j=i+1}^{\numaminoacids} g(d_{ij}),
\end{equation}
where the indicator function $g(d_{ij}) = 1$ if the distance $d_{ij} = 0$, and $g(d_{ij}) = 0$ otherwise.
Up to this point, we have made sure that any physically forbidden configurations, due to either overlapping beads or redundant encodings, are properly penalized in the energy function, so that they do not appear in the low-energy region of the solution space.

Finally, we consider the interaction energy term, which captures the pairwise interactions between amino acids based on their distances.
The interaction term is what differentiates among the physically allowed conformations and is responsible for guiding the optimization process toward the ground state, which is assumed to be the approximate solution to the \gls{psp} problem.
Due to the coarse-grained nature of the model, the interaction energy is typically defined on the amino acid level, rather than the atomistic level.
The simplest interaction model is the \gls{hp} model~\cite{lau_lattice_1989}, which classifies amino acids as either hydrophobic or polar.
It then assigns a negative energy ($-1$) only if two hydrophobic amino acids are in contact, i.e., \knn apart, and zero otherwise.
This model captures the phenomenon of hydrophobic collapse, where hydrophobic residues tend to cluster together in the core of a protein structure.
A more sophisticated interaction model is the \gls{mj} model~\cite{miyazawa_residue_1996}, which assigns different interaction energies to different pairs of amino acids depending on their chemical properties.
Roughly speaking, the statistical potential in the \gls{mj} model is derived from observed frequencies of amino acid contacts in experimentally determined protein structures.
In this work, we use the \gls{mj} energy model, which is defined by a symmetric $20\times 20$ interaction matrix, where each entry corresponds to the interaction energy between the two unique amino acids.
Therefore, the energy term for the \knn interactions can be expressed as:
\begin{equation}
    E_{\text{int}}^{(k)}(\vb q) = \sum_{i=1}^{\numaminoacids - 1} \sum_{j=i+1}^{\numaminoacids} \frac{\varepsilon_{ij} \cdot h^{(k)}(d_{ij})}{d_{ij} / d^{(1)}},
\end{equation}
where $\varepsilon_{ij}$ is the interaction energy between amino acids $i$ and $j$, and $h^{(k)}(d_{ij})$ is an indicator function that is equal to 1 if the distance $d_{ij}$ is equal to the \knn distance for a given lattice, and zero otherwise.
For example, on the tetrahedral lattice, the 1-NN distance is $d^{(1)}_\text{tetra} = \sqrt{3}$, which can be computed as the norm of any turn basis vector shown in Table~\ref{tab:tetra_turns}.
The 2-NN distance is $d^{(2)}_\text{tetra} = \sqrt{8}$, the 3-NN distance is $d^{(3)}_\text{tetra} = \sqrt{11}$, and so on.
Similarly, for the \gls{bcc} lattice, we have $d^{(1)}_\text{bcc} = \sqrt{3}$, $d^{(2)}_\text{bcc} = 2$, $d^{(3)}_\text{bcc} = \sqrt{8}$, and so on.
For the \gls{fcc} lattice, we have $d^{(1)}_\text{fcc} = \sqrt{2}$, $d^{(2)}_\text{fcc} = 2$, $d^{(3)}_\text{fcc} = \sqrt{6}$, etc.
Note that the interaction energy for the \knn beads is scaled by the ratio of the \knn distance to the 1-NN distance, $d_{ij} / d^{(1)}$, to reflect the Coulomb-like decay of the interaction energy with distance.
The total interaction energy is then obtained by summing over all \knn interactions:
\begin{equation}
    E_{\text{int}}(\vb q) = \sum_{k=1}^{\maxkNN} E_{\text{int}}^{(k)}(\vb q),
\end{equation}
where $\maxkNN$ is the maximum number of nearest neighbors considered in the model.

\subsection{Optimization details \label{sec:optimization}}

The experiments presented in this work consist of two main steps: training the \gls{hea} parameters on a noiseless simulator and sampling the optimized circuit on a simulator or a real quantum device.
In the first step, we use a statevector simulator that performs exact quantum circuit simulations for up to 30 qubits, and an \gls{mps} simulator for up to 46 qubits.
The statevector simulation, while providing exact results, incurs a computational cost that exponentially increases with the number of qubits.
The \gls{mps} simulator, on the other hand, enables efficient simulations of larger systems by approximating the quantum state with a tensor network of constant bond dimension.
In our experiments, we focus mainly on benchmarking the performance of the \gls{hea} with a single layer and up to 46 qubits.
The relatively shallow depth of these circuits allows for an accurate \gls{mps} representation with a moderate bond dimension, making it feasible to simulate larger protein instances.

The optimization of \gls{hea} parameters is performed using the COBYLA algorithm~\cite{cobyla_Powell1994}, which is a derivative-free optimization method that is more suitable for complex and noisy cost functions.
The maximum number of iterations is set to 5,000 throughout the experiments. 
However, the algorithm typically terminates before reaching this limit when it fails to detect any improvement for several consecutive steps, especially for smaller proteins.
Additionally, we adopt the Conditional Value-at-Risk (CVaR) optimization strategy~\cite{barkoutsos_improving_2020}, where the cost function to be minimized is defined as the average of the low-energy tail of the energy distribution (computed based on the energy function defined in Sec.~\ref{sec:energy_function}), delimited by a threshold value $\alpha$.
The CVaR cost function has been shown to provide a faster convergence to low-energy states compared to the standard cost function based on the entire energy distribution.
We set the threshold value $\alpha$ to 0.1, meaning that we only consider the lowest 10\% of the measured energies at each optimization step.
Furthermore, to partially mitigate stochasticity, we repeat the optimization for each protein candidate with ten different random initializations of the circuit parameters.
Within each run, we record all measured bitstrings throughout the optimization process and rank them based on their corresponding energies at the end.
It allows us to select the best bitstrings---those with the lowest energies---recorded during training, regardless of whether they were measured at the end of the optimization or an earlier stage, for further analysis and potential post-processing.
This differs from the standard practice in quantum \gls{psp}, where the best solutions are typically determined based solely on the final measurement results.
In Sec.~\ref{sec:results}, we present both the average quantities over the ten runs and the corresponding metrics based on the best measured bitstring; more details are provided in Sec.~\ref{sec:metrics}.

In the second step, we sample the optimized quantum circuit on a noiseless simulator or a real quantum device. 
It is worth noting that no further training is performed at this stage; the circuit parameters are already trained in a fully classical environment.
The sampled bitstrings and their corresponding energies are then used to evaluate the performance metrics described in Sec.~\ref{sec:metrics}, providing insights into the performance of the quantum circuit with and without noise.

\subsection{Performance metrics}
\label{sec:metrics}

To evaluate the performance of our quantum \gls{psp} workflow, we employ a set of complementary metrics, each offering unique insights into the quality of the solutions produced. 
The output of the quantum circuit is consistently evaluated in relation to the energy of the lowest-energy structure, known as the ground state energy, denoted by $\groundstateenergy$. 
It is worth noting that even if the exact ground state is not reached, proximity to it may make it easier to achieve the ground state with the help of post-processing, such as simulated annealing~\cite{kirkpatrick1983optimization}.
The ground state energies for the proteins investigated in this study are determined through an exhaustive search of the entire solution space. 
This exhaustive search method generates all possible self-avoiding walks of a given number of steps (i.e., the number of turns in a protein chain) on a lattice and sorts them by energy in real time.
However, the computational time scales exponentially with system size and, therefore, this approach is only feasible for the sizes of proteins considered in this work.

To begin with, we visualize the probability of measuring bitstrings with different energies by plotting the \textit{energy distribution} of the bitstrings sampled from the optimized quantum circuit.
This distribution provides a qualitative view of the ansatz's performance, highlighting the likelihood of it sampling low-energy configurations, particularly those near the ground state.
It is important to note that the energy distribution can also reveal potential gaps in the optimization process, such as a lack of exploration in certain energy regions.
It is also helpful to compare the energy distributions obtained from different lattice types or different interaction orders, as this can shed light on how these factors influence the optimization landscape and the circuit's ability to find low-energy states.
To enable such a comparison across various protein instances, we normalize the energy values by the absolute value of the ground state energy for all proteins.

Further, to quantify more precisely how close the sampled solutions' energies are to the exact ground state energy, we use the \textit{relative error (RE)}.
This metric is widely used in quantum optimization, as it provides a straightforward measure of how well the quantum circuit can generate approximately optimal solutions. 
Relative error is similar to the approximation ratio, which is typically defined as a simple ratio of the algorithm's cost to the optimal cost of a problem.
However, relative error is more suitable than approximation error for minimization problems, especially in our case, where the optimal energy of a protein is always negative. In contrast, the energies of other suboptimal conformers can deviate significantly from the optimum, or even be positive. 
While the approximation ratio can still provide insight, it becomes less suitable and potentially deceitful in our context, where a positive energy, far from the optimal value, is translated to a negative approximation ratio.
In contrast, relative error is bounded by one and remains a stable and interpretable metric across our problem instances, offering a more robust basis for comparison.
A relative error near zero implies that the quantum circuit produced a low-energy state whose energy is close to the ground state energy.

In our analysis, we use the relative error in two ways.
First, we calculate the relative error using the average cost of the sampled bitstrings from the quantum circuit after optimizing its parameters.
It measures the average performance of the trained quantum circuit in generating the low-energy ensemble of conformers for a protein.
We call this \gls{average_error}, defined as
\begin{equation} \label{eq:average_error}
    \text{\gls{average_error}} = \abs{ \frac{C_{\text{CVaR}} - \groundstateenergy}{\groundstateenergy} },
\end{equation}
where $C_{\text{CVaR}}$ represents the average energy of the lowest 10\% of the sampled conformers from the optimized quantum circuit, and $\groundstateenergy$ denotes the ground state energy of the protein instance.

Second, we compute the relative error with respect to the lowest energy from Eq.~\eqref{Eq:energy_func} observed throughout the entire training process, providing a measure of the best performance achieved by the method.
We call this \gls{best_case_error}, defined as
\begin{equation} \label{eq:best_case_error}
    \text{\gls{best_case_error}} = \abs{ \frac{E_{\text{lowest}} - \groundstateenergy}{\groundstateenergy} },
\end{equation}
where $E_{\text{lowest}}$ is the lowest energy measured during the whole training.
This metric is particularly important for the \gls{psp} problem, where the goal is to find the lowest-energy conformation of a protein.
While near-optimal solutions may suffice in some optimization problems, in \gls{psp}, these solutions can be significantly less useful, as a slight difference in the structure can lead to much different biochemical properties.

Together, \gls{average_error} and \gls{best_case_error} offer complementary insights into the performance of the quantum workflow. 
The \gls{average_error} reflects the effectiveness of the final circuit by summarizing the overall energy landscape sampled after optimization. 
In contrast, the \gls{best_case_error} measures if the ground state is found and if not, how close the best sampled configuration approaches the ground state energy throughout the entire workflow.
In hardware experiments, we do not perform any further optimization of the circuit parameters; instead, quantum circuits with the trained parameters are measured with 100,000 shots, and the sampled bitstrings are recorded to compute the aforementioned metrics. 
As a result, the \gls{best_case_error} from the hardware runs indicates the closest solution to the ground state that the quantum device can produce, while the \gls{average_error} from the hardware runs serves as a statistical summary of the sampled energy distribution.
For comparison, we also sample the optimized circuits on a noiseless simulator with the same number of shots (100,000) and compute the same metrics.
This allows us to assess the impact of noise and other hardware-specific factors on the performance of the quantum circuits.
Note that we do not report the commonly used \textit{hit rate}, i.e., how many times the ground state is measured, as the exact ground state was not observed in any of the larger protein instances, making this metric less informative in our context.


\section{Results}
\label{sec:results}

We investigate the effectiveness of the Hamiltonian-free \gls{psp} workflow utilizing \glspl{hea} in predicting low-energy protein structures for a set of proteins across different lattice encodings---tetrahedral, \gls{bcc}, and \gls{fcc}---for both 1-NN and 2-NN interaction models. 
As mentioned earlier, after training the quantum circuits on a noiseless simulator, we sample the optimized circuits on both a noiseless simulator and a real quantum device to evaluate their performance.
Due to the inconsistent noise profiles of the quantum device over time, we performed five independent samplings with 100,000 shots for every protein instance over the course of two days.
The quantum hardware experiments were conducted on the \texttt{ibm\_kingston} device, which is a Heron R2 quantum processor developed by IBM.
The hardware experiments on the IBM device were conducted both with and without error suppression techniques, but no substantial improvements were observed in the results. 
In the following, we present the results obtained with error suppression techniques, which include Pauli twirling~\cite{wallman2016noise} and dynamical decoupling~\cite{ezzell2023dynamical} with the XX pulse sequence.

We evaluate a set of problem instances described in Appendix~\ref{app:protein_candidates} (see Table~\ref{tab:protein_instances}) using the performance metrics introduced in Sec.~\ref{sec:metrics}.
Each protein instance is denoted by its Protein Data Bank Identification Code (PDB-ID)~\cite{burley2025updated}. 
For consistency, the 7-amino-acid Alpha-Bag Cell Peptide~\cite{rothman1983primary}, which does not have a PDB-ID, is abbreviated as ``ABCP''.
The proteins were selected to cover a range of lengths (from 5 to 26 amino acids) and structural complexities, including secondary structure elements such as alpha-helices and beta-sheets.
The largest protein instances for each lattice type correspond to more than 40 qubits, beyond which the classical exhaustive search for the ground state becomes infeasible due to the exponential growth of the solution space.
In the following subsections, we present the simulator and quantum hardware results of these proteins, focusing on the energy distributions, \gls{average_error}, and \gls{best_case_error}.

\subsection{Energy distributions}
\label{sec:prob_distr}

\begin{figure*}[ht]
    \centering
    \includegraphics[width=1.0\linewidth]{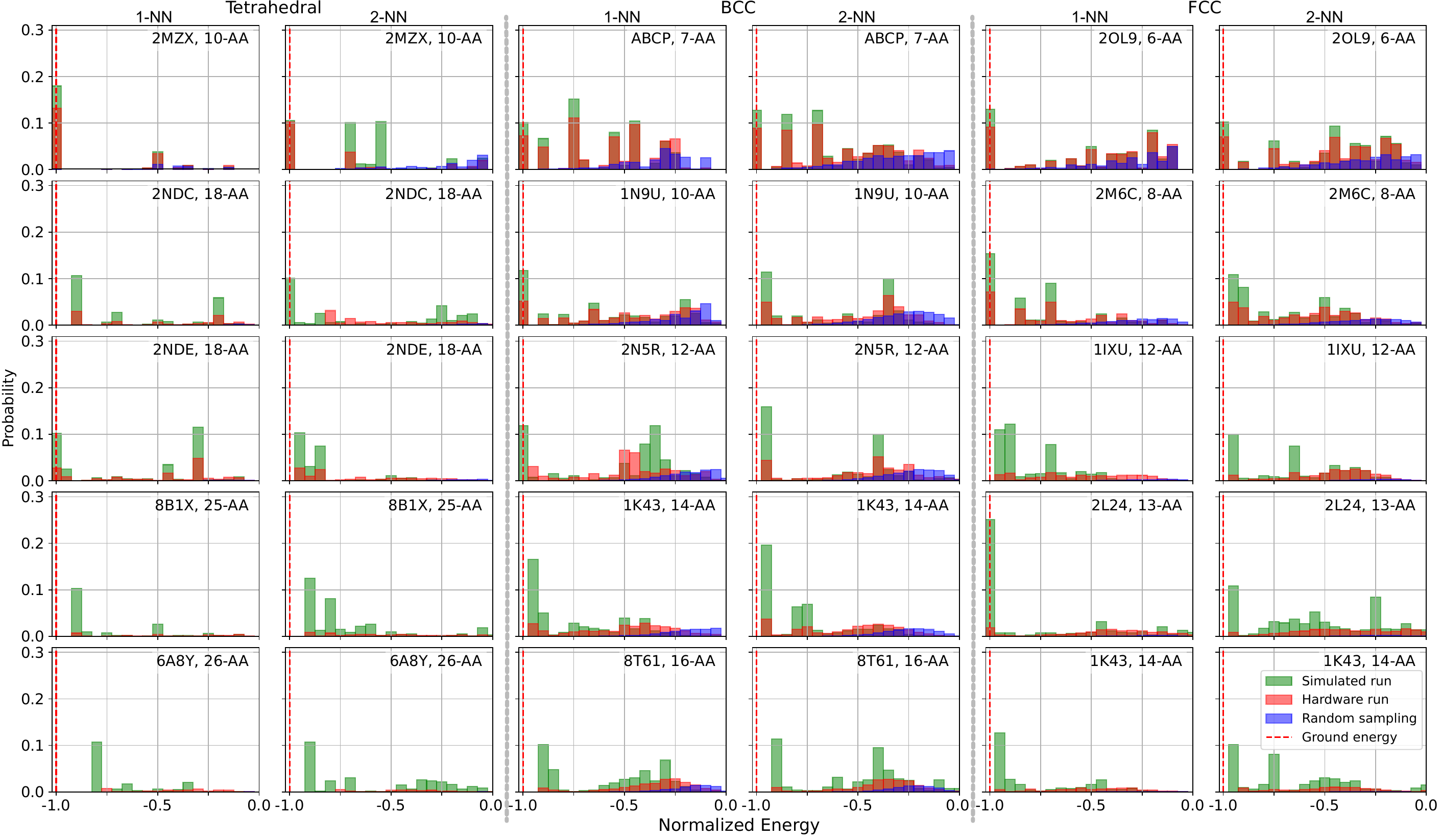}
    \caption{\textbf{Probability distribution of sampled energies from the best optimized quantum circuit.} The energies are normalized with respect to the ground state energy, indicated by red dashed lines. The y-axis represents the probability of measuring a given state, while the x-axis shows its energy relative to the ground state energy of the protein. Noiseless simulation results are shown in green, and hardware results from \texttt{ibm\_kingston} are shown as red histogram bins. For reference, the random sampling baseline with 100,000 bitstrings is shown in blue. Results are grouped by lattice encoding---tetrahedral, \gls{bcc}, and \gls{fcc}---and include both 1-NN and 2-NN interaction models.
    Alpha-Bag Cell Peptide is abbreviated as ``ABCP'' to maintain consistency with other names.}
    \label{fig:prob_distr}
\end{figure*}

The data presented in Fig.~\ref{fig:prob_distr} is the probability distributions of energies of the measured bitstring, each corresponding to a potential protein conformation, measured from the optimized quantum circuit.
For each protein instance, the distribution is generated from the quantum circuit with the lowest CVaR cost out of ten optimization runs on the noiseless simulator and out of five sampling runs on the quantum hardware.
It is worth noting that we only present the bitstrings that correspond to physically valid conformations, which have negative energies.
The figure visualizes the probability of sampling states with energies close to the ground state for each amino acid sequence represented by its PDB-ID, across different lattice encodings---tetrahedral, \gls{bcc}, and \gls{fcc}---and for both 1-NN and 2-NN interaction models. 
Only a subset of the studied protein instances is shown; the complete set of results is available in Appendix~\ref{app:complete_results}.
In Fig.~\ref{fig:prob_distr}, the x-axis shows the normalized energy, defined as the energy of a state divided by the absolute value of the ground state energy, with the ground state energy indicated by red dashed lines.
The y-axis represents the probability of energy windows sampled from the optimized circuit.
The noiseless simulation results are shown in green, and hardware results from \texttt{ibm\_kingston} are shown as red histogram bins.
Additionally, for each protein instance, we also include a baseline distribution obtained from randomly sampling 100,000 bitstrings, shown in blue.
This baseline provides a point of comparison to assess the effectiveness of the quantum optimization process.

In the noiseless simulations, the probability of measuring a state close to the ground state is generally below 20\%, and occasionally zero, with one exception of 2L24 in the \gls{fcc} lattice with 1-NN interactions that goes to almost 25\%.
A general trend across all lattice types and interaction orders is that as the protein size increases, the lowest-energy bin in the histogram tends to shift further away from the ground state energy.
However, even when the ground state is not sampled directly, the distributions often show significant probability one or a few steps away in the normalized energy from the ground state energy, which may be sufficient for classical postprocessing techniques to recover the optimal structure. 
In most cases, despite the distributions typically showing significant mass away from the ground state, the lowest-energy bin is often dominant, showing the highest probability among all bins.
In comparison, random sampling generally results in a much flatter distribution, with a very low, often negligible, probability of sampling states near the ground state.
As the protein size increases, the random sampling distribution becomes less noticeable in the histograms of negative energies, indicating that the solution space is becoming increasingly sparse relative to the number of samples taken.
This means that as proteins get larger, randomly guessing conformations is less and less likely to yield low-energy states simply due to the vastness of the possible conformation space.
In contrast, the \gls{hea} is able to concentrate the probability distribution around low-energy states, even if it does not always reach the ground state itself.

There is no general trend across the different lattice encodings on which protein instances are most challenging to optimize. 
For example, the longest sequence in the tetrahedral encoding of the 26-amino-acid protein 6A8Y yields a distribution far from the ground state, which aligns with its high circuit resource demand---45 qubits and 134 gates. 
In contrast, the largest \gls{fcc} instance, 1K43---which requires a slightly larger circuit with 46 qubits and 137 gates---achieves a distribution with a high probability of sampling near the ground state.
The \gls{bcc} encoding of 8T61, using 41 qubits, also shows limited success in sampling low-energy states, though not as severely as in the tetrahedral case.
These contrasting outcomes suggest that optimization difficulty is not solely determined by circuit size; this is likely due to differences in the energy landscapes. 
Nevertheless, all larger instances---regardless of encoding---tend to be challenging, with broader distributions and reduced probability of sampling the ground state.
Similarly, the comparison between 1-NN and 2-NN models reveals mixed results. In some cases, 
2-NN plots show performance by increasing the probability of sampling low-energy states, e.g., tetrahedral instances 2NDC, 8B1X, and 6A8Y; \gls{bcc} instances ABCP and 1K43.
While in others, 1-NN yields better outcomes, e.g., tetrahedral instance 2MZX; \gls{bcc} instances 1N9U and 2N5R; \gls{fcc} instances 2M6C, 2L24, and 1K43.

Experimental results obtained from quantum hardware typically exhibit broader probability distributions and reduced sampling near the ground state relative to the noiseless simulations. 
As anticipated, shorter amino acid sequences---corresponding to shallower quantum circuits with fewer qubits---yield distributions more closely aligned with the ground state, whereas larger instances exhibit increased divergence due to increased circuit depth and qubit count.
This deviation reflects the influence of noise and imperfections inherent in current quantum devices. 
Nevertheless, qualitative trends observed in simulation are often preserved in hardware outcomes. 
For instance, in the case of the ABCP protein mapped onto the \gls{bcc} lattice, both the 1-NN and 2-NN configurations show that states with high sampling frequency in simulation are similarly favored in hardware runs. 
Moreover, the hardware results still outperform random sampling baselines, with higher probabilities of sampling low-energy states across all protein instances.
This implies that even for the largest instances studied, the quantum circuits retain some ability to generate low-energy conformers despite the presence of noise.

\subsection{Average relative error}
\label{sec:are}

\begin{figure*}[!ht]
    \centering
    \includegraphics[width=0.90\linewidth]{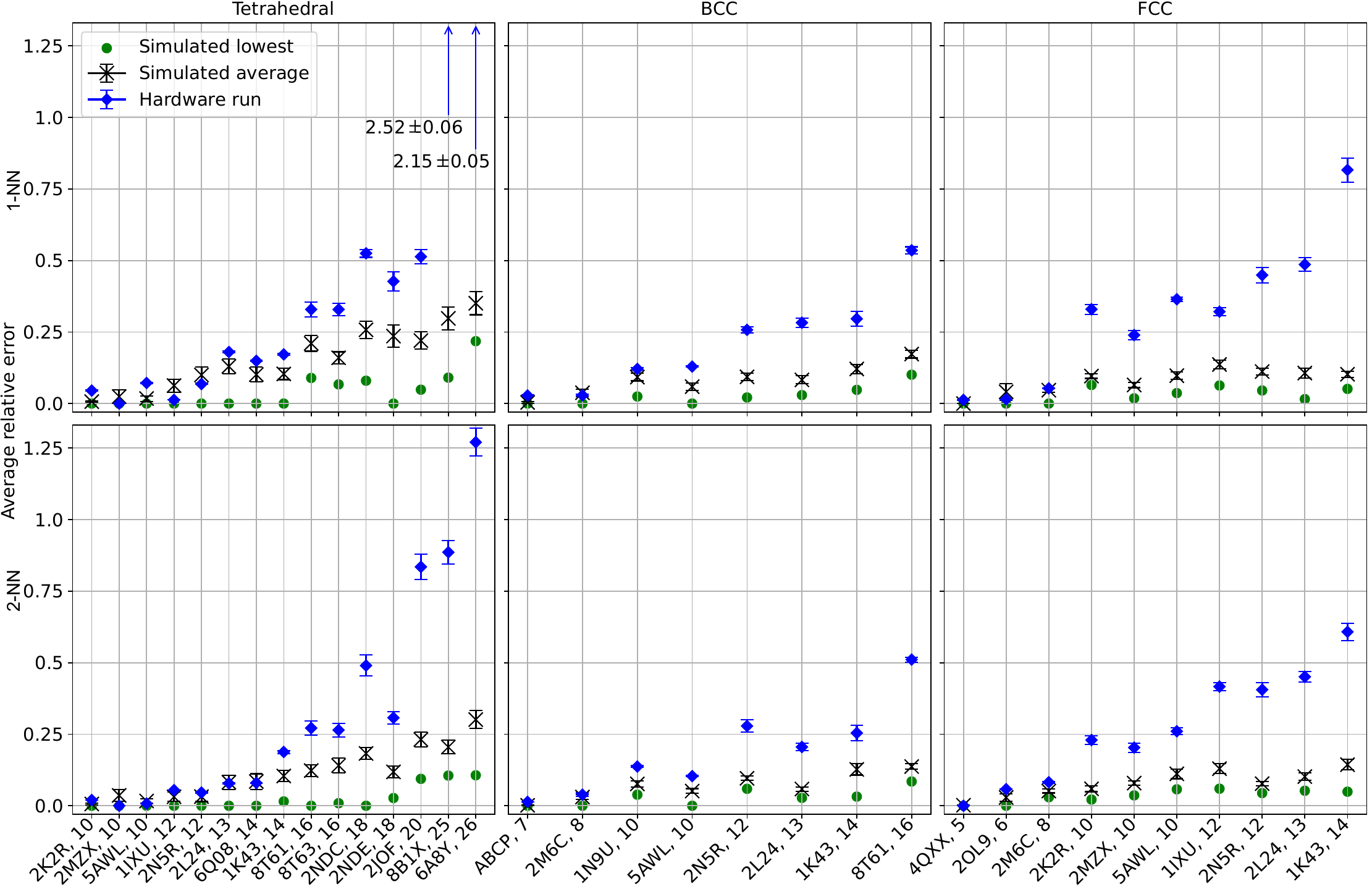}
    \caption{\textbf{Average relative error computed using the CVaR cost function, for sampled bitstrings obtained after circuit parameter optimization.} 
    The x-axis lists the protein instances considered, annotated with their respective number of amino acids. 
    The y-axis shows the \gls{average_error}.
    Black crosses denote \glspl{average_error} from classical simulations---statevector simulations for circuits up to 30 qubits and \gls{mps} simulations for larger circuits---each averaged over the ten optimization runs, with the standard error shown.
    Green dots indicate the lowest error of those ten runs.
    Blue diamonds represent results obtained from sampling the optimized circuits on the \texttt{ibm\_kingston} quantum device, averaged over five independent runs.}
    \label{fig:average_error}
\end{figure*}

Next, we evaluate the \acrlong{average_error}, as defined in Eq.~\eqref{eq:average_error}, to assess the performance of the optimized quantum circuits in generating low-energy ensembles of conformers for each protein instance.
The results are presented in Fig.~\ref{fig:average_error}, where the x-axis lists the protein instances by their PDB-ID, annotated with the number of amino acids, and the y-axis shows the \gls{average_error}, indicating how far the CVaR-average cost of the sampled conformers is from the ground state energy.
Each column (row) corresponds to a specific lattice encoding (interaction order), with results grouped accordingly.
In each subplot, the black crosses indicate results from classical noiseless simulations. 
These include statevector simulations for circuits up to 30 qubits and \gls{mps} simulations for circuits beyond that. 
For each protein instance, the results are averaged over ten different optimization runs, and the corresponding standard errors are shown.
The green dots represent the lowest average cost in those ten simulator runs.
Similarly, the blue diamonds denote the CVaR-average cost with the same threshold ($\alpha = 0.1$) obtained from sampling the optimized circuits on the \texttt{ibm\_kingston} quantum device, with each data point corresponding to the average of five independent hardware runs and the error bars indicating the standard error.

First, focusing on the simulator results, we observe that the \gls{average_error} shown in black crosses generally increases with the size of the protein instance, or equivalently, the number of qubits in the circuit.
This trend is expected, as larger proteins present more complex energy landscapes and require more parameters to optimize, making it more challenging for the \gls{hea} to effectively explore the solution space.
However, there are exceptions to this trend, such as the tetrahedral encoding of 2NDE with 2-NN interactions, which shows a lower \gls{average_error} than some smaller instances.
This suggests that factors beyond just the number of qubits, such as the specific energy landscape of the protein, determined by its sequence and the chosen energy matrix, also play a significant role in the optimization difficulty.
Lattice type also influences performance, with the tetrahedral encoding generally yielding higher \gls{average_error} values compared to \gls{bcc} and \gls{fcc} encodings for instances with similar numbers of qubits.
For example, comparing the largest instances on the tetrahedral lattice, 6A8Y with 45 qubits, and on the \gls{fcc} lattice, 1K43 with 46 qubits, we see that the tetrahedral instance has a significantly higher \gls{average_error} for both 1-NN and 2-NN interactions.
This is consistent with what is observed in the energy distributions of the two, as reported in Sec.~\ref{sec:prob_distr}.
Regarding the interaction order, the results are mixed, but overall, the performance does not show a consistent trend favoring either the 1-NN or 2-NN models across different instances.
However, it should be noted that this observation is based on the fact that the reference energy, i.e., ground state energy of the coarse-grained model, changes when moving from 1-NN to 2-NN models, making direct comparisons of \gls{average_error} between the two interaction orders less straightforward.
In practical \gls{psp} applications, one would typically compare the predicted structures to experimentally determined ones, rather than relying solely on the coarse-grained model's ground state energy.
In that case, the inclusion of higher-order interactions would likely lead to better structural predictions because they capture more realistic physical interactions.
Nevertheless, the lowest \gls{average_error} values, indicated by the green dots, are often significantly lower than the average values, with the worst ARE being around 20\% for the largest 1-NN tetrahedral instance, 6A8Y. 
This suggests that the optimizer can sometimes find parameter settings that yield much better performance.
In these best-case scenarios, the performance appears less sensitive to lattice type and interaction order, indicating that with sufficient optimization effort, the \gls{hea} can effectively learn to generate low-energy ensembles for various configurations, achieving an \gls{average_error} below 10\% in the largest instance with 46 qubits.

Turning to the hardware results, shown as blue diamonds in Fig.~\ref{fig:average_error}, we observe that the \gls{average_error} values are generally higher than those obtained from noiseless simulations, reflecting the impact of noise and other imperfections in the quantum device.
This trend is particularly pronounced for larger protein instances, where the hardware \gls{average_error} can be several times higher than the simulator results.
For example, in the largest 1-NN instances of each lattice type, 6A8Y for tetrahedral, 8T61 for \gls{bcc}, and 1K43 for \gls{fcc}, the hardware \glspl{average_error} are roughly three to eight times those of the simulator results.
Such more drastic performance degradation compared to the simulator results is due to the fact that the circuit depth, and more importantly, the two-qubit gate depth, increases linearly with the number of qubits in the \gls{hea} used here (cf.~Fig.~\ref{fig:scheme}).
As a result, larger protein instances require not only more qubits but also deeper circuits, which are more susceptible to noise and decoherence effects in current quantum hardware.
On the other hand, for smaller protein instances on each lattice, the hardware \gls{average_error} values are often comparable to the simulator results.
Together with the tight error bars, this demonstrates the robustness and consistency of the \texttt{ibm\_kingston} device for these smaller and shallower circuits with around 25 qubits or fewer.
Overall, the hardware results highlight the challenges of scaling the current \gls{hea} to larger protein candidates on noisy quantum devices, while also showing promise for smaller instances where the hardware can achieve performance close to that of noiseless simulations.

\subsection{Best-case relative error}
\label{sec:best_case_error}

\begin{figure*}[!ht]
    \centering
    \includegraphics[width=0.90\linewidth]{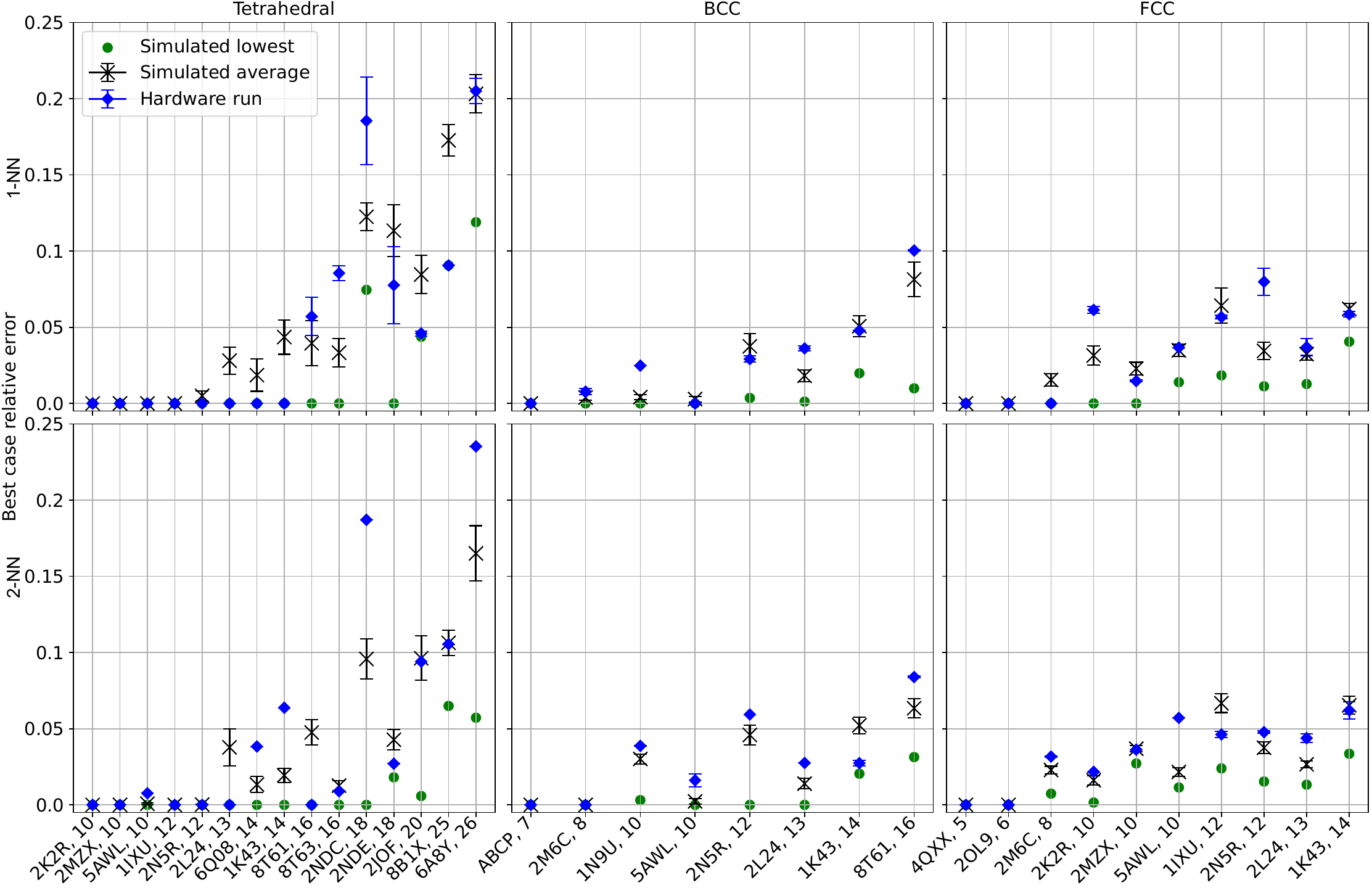}
    \caption{\textbf{Best-case relative error, defined as the relative error of the lowest energy observed throughout the entire training process.}
    The x-axis lists the protein instances considered, annotated with their respective number of amino acids.
    The y-axis shows the \gls{best_case_error}.
    Black crosses denote the \glspl{best_case_error} from the simulator runs, each averaged over ten runs, with standard error shown. 
    Green dots indicate the lowest \gls{best_case_error} of those ten runs.
    Blue diamonds represent results obtained from sampling the optimized circuits on the \texttt{ibm\_kingston} quantum device, averaged over five independent runs.}
    \label{fig:best_case_error}
\end{figure*}

To assess the potential of the full Hamiltonian-free \gls{psp} workflow in finding the lowest-energy conformation close to the ground state(s) of a protein, we examine the \acrlong{best_case_error}, as defined in Eq.~\eqref{eq:best_case_error}.
The simulator and hardware results are presented in Fig.~\ref{fig:best_case_error}, in a similar format to Fig.~\ref{fig:average_error} for the \acrlong{average_error}.
In this case, the best-case energy recorded in simulations is defined as the lowest energy observed throughout the entire training process of a quantum circuit.
For hardware experiments, it is the lowest energy measured from sampling the optimized circuit with the trained parameters, as no optimization is performed on the quantum device.
As mentioned in the previous subsection, each of the simulated averages (black crosses) is the mean of ten simulator runs with random initial parameters, whereas the hardware averages (blue diamonds) are the mean of five independent hardware runs.
The green dots indicate the lowest \gls{best_case_error} in those ten simulator runs.

Comparing with the \glspl{average_error} obtained on the simulator, as shown in Fig.~\ref{fig:average_error}, one immediately observes that the \glspl{best_case_error} from simulations (black crosses) are generally lower across all protein instances.
Such differences are also reflected in the lowest \gls{best_case_error} values (green dots), which are often lower than the corresponding \gls{average_error} values for the same protein instance.
For example, all the tetrahedral proteins with lengths up to 16 amino acids achieve a lowest \gls{best_case_error} that is nearly zero, indicating that the ground state was sampled at least once during at least one of the ten optimization runs.
For smaller instances, e.g., below 12 amino acids on the tetrahedral lattice, even the average \glspl{best_case_error} are nearly zero, suggesting that the \gls{hea} can consistently find the ground state or states very close to it regardless of the random initialization.
Similar trends are observed for the \gls{bcc} and \gls{fcc} lattices, where the \glspl{best_case_error} are consistently close to zero for smaller proteins, and the lowest \glspl{best_case_error} remain below 5\% even for the largest instances with up to 46 qubits.

On the hardware side, the \glspl{best_case_error} (blue diamonds) are also generally lower than the corresponding \glspl{average_error} shown in Fig.~\ref{fig:average_error}, and the differences are even more pronounced.
For instance, the largest tetrahedral instance, 6A8Y, achieves a \gls{best_case_error} of around 20\%, which is significantly lower than its \gls{average_error} of around 215\%.
A logical explanation for this is that the hardware noise can degrade the overall quality of the sampled energy distribution, leading to more erroneous samples that might have very high energies.
Due to the nature of the turn-based encoding, a single bit flip in the solution bitstring can lead to a large change in the overall conformation of a protein, and consequently, a large change in its energy, especially if the new conformation contains geometric violations such as bead overlaps.
As a result, the average energy of the sampled conformers can be significantly affected by these high-energy outliers, leading to a high \gls{average_error}. Such an effect can be mitigated by using a CVaR cost function with a lower threshold value $\alpha$, which focuses more on the low-energy tail of the distribution.
In contrast, the \gls{best_case_error} is determined solely by the lowest energy sampled, which is less likely to be affected by these outliers.
Therefore, even in the presence of hardware noise, the \gls{best_case_error} can still provide a reliable estimate of the circuit's ability to find the ground state or states close to it.
Overall, the \gls{best_case_error} results demonstrate the potential of the proposed \gls{psp} workflow in finding low-energy conformations close to the ground state, even on noisy quantum hardware.

\section{\label{sec:discussion} Discussion and outlook}

In summary, we have demonstrated a scalable quantum-classical hybrid workflow for protein structure prediction that leverages problem-agnostic ansatzes, such as a hardware-efficient one. 
By offloading structural constraints and energy evaluations to classical post-processing, our method avoids the need for explicit Hamiltonian construction and encoding of protein-specific information into the quantum circuit, which typically leads to excessive ancilla qubit overhead and deep circuits, as seen in previous methods~\cite{babbush2014construction, babej_coarse-grained_2018, robert_resource-efficient_2021, liQuantumAlgorithmProtein2025}.
As a result, the qubit requirements are significantly reduced, scaling linearly with the number of amino acids, regardless of the lattice type or the order of interactions considered.
In comparison, the best Hamiltonian-based methods to date scale quadratically with just 1-NN interactions~\cite{robert_resource-efficient_2021}.
Moreover, it circumvents the difficulty of encoding geometric constraints on a lattice and higher-order interactions into a Hamiltonian, which can be challenging and often requires additional qubits.
In our approach, higher-order interactions can be easily incorporated into the energy function with little classical overhead, making it more flexible and adaptable to different lattice types and interaction models.
Unlike earlier proof-of-concept studies, which were limited to small peptides with a few amino acids, we have benchmarked our Hamiltonian-free workflow on a comprehensive set of proteins up to 26 amino acids, using three different lattice types and up to 2-NN interactions.
This marks the largest protein instances run on quantum hardware to date, showcasing the scalability of our approach.
These experiments push the current quantum systems to their limits, with the largest circuit consisting of 46 qubits, 45 two-qubit gates, and a depth of 143.
Moreover, extending to higher-order interactions is straightforward and computationally efficient in our workflow.

The simulator results demonstrate that the \gls{hea} with a single layer successfully learned to generate low-energy ensembles for most of the protein instances tested in this work, with peaks close to the ground state shown in the energy distributions.
Quantitatively, the \acrlong{average_error} from the simulator runs generally increases with the size of proteins, yet the lowest \gls{average_error} values achieved in the ten runs remain below 25\% for all instances, with the largest instance with 46 qubits achieving an \gls{average_error} below 10\%.
Moreover, our results show that the workflow can effectively find the ground state for smaller proteins, with the lowest \acrlong{best_case_error} being nearly zero for all instances with up to 16 amino acids on the tetrahedral lattice, 13 amino acids on the \gls{bcc} lattice, and 10 amino acids on the \gls{fcc} lattice.
For larger proteins, the lowest \gls{best_case_error} values remain below 13\% for all instances, with the largest instance (46 qubits) achieving a lowest \gls{best_case_error} below 5\%.
Compared across different lattice types and interaction orders, an interesting observation is that the performance of the \gls{hea} appears to be more sensitive to the lattice type than the interaction order, with the tetrahedral encoding generally yielding higher relative errors compared to \gls{bcc} and \gls{fcc} encodings for instances with similar numbers of qubits.
We suspect that this is due to the more complex energy landscape associated with the tetrahedral lattice, which has a lower coordination number and therefore more geometric constraints when compared to the other two encodings under the same number of qubits, making it harder for the variational quantum circuit to effectively explore the solution space.
On the other hand, the hardware results obtained on \texttt{ibm\_kingston} show that the \gls{hea} can still generate meaningful structures for smaller proteins, with relative errors being comparable to those from noiseless simulations for instances that correspond to around 25 qubits or fewer.
However, the performance degrades significantly for larger proteins, due to the increased number of qubits, and more importantly, the deeper circuits required, which are more susceptible to noise effects in current quantum hardware.
This calls for improved hardware with lower noise rates together with more efficient ansatz designs to fully realize the potential of quantum computing for large-scale protein structure prediction.
Nevertheless, compared to random sampling, the hardware results still outperform it even in the largest instances, showing discernibly higher probabilities of sampling low-energy physical states, as seen in Fig.~\ref{fig:prob_distr}.

As mentioned earlier, a significant strength of our approach is its scalability. 
In this work, the use of an \gls{hea} together with the Hamiltonian-free classical post-processing ensures that the number of qubits and gates scales linearly with the size of proteins, making it more practical for near-term quantum devices. 
However, the expressibility of the one-layer \gls{hea} may be limited for larger proteins, as it may not be able to capture the complex energy landscape of protein folding with a shallow circuit.
Additionally, \glspl{hea} are known to exhibit barren plateaus in the parameter landscape when scaled to larger systems and deeper circuits, which may pose a significant challenge to optimization~\cite{mcclean2018barren, larocca2025barren}.
These challenges are partially reflected in the noiseless simulator results, where we observe a degradation in performance as the protein size increases across all three lattices.
Moreover, preliminary simulation results with the two-layer \gls{hea} on the tetrahedral lattice (not shown) did not show noticeable improvements over the one-layer ansatz, suggesting that deeper circuits with more parameters may be harder to train due to the increased complexity in optimization.
Further investigations into the expressibility and trainability of the \glspl{hea} for larger proteins are warranted to fully understand their potential and limitations.

One of the future directions is to explore more advanced ansatz designs that can further improve the accuracy of the predicted structures for larger proteins, while still maintaining the trainability and scalability on near-term quantum devices.
It is worth emphasizing that the proposed workflow is not limited to \glspl{hea} and can be adapted to other problem-agnostic ansatzes, such as the instantaneous quantum polynomial (IQP) ansatz~\cite{shepherd2009temporally, havlicek2019supervised}, various tensor-network inspired ansatzes~\cite{huggins2019towards, du2020expressive, haghshenas2022variational}, adaptive ansatzes~\cite{grimsley2019adaptive,tang2021qubit}, etc., which may offer better trainability and other advantages for larger systems.
Furthermore, numerous strategies have been proposed to mitigate barren plateaus, such as layer-wise training~\cite{skolik2021layerwise}, local cost functions~\cite{cerezo2021cost}, and better parameter initialization schemes~\cite{grant2019initialization, park2024hardware}.
Another promising direction for future work is the inclusion of higher-order interactions beyond 2-NN. 
These interactions can capture more subtle and long-range effects in protein folding, potentially leading to more accurate predictions.
In our Hamiltonian-free workflow, higher-order interactions can be easily incorporated into the energy function with minimal additional overhead.
This flexibility allows one to explore more sophisticated interaction models, such as those based on machine learning potentials~\cite{majewski2023machine}, which can learn complex interaction patterns from large datasets of protein structures.
It is worth exploring alternative energy matrices beyond the \gls{mj} model used here, which may offer improved biological relevance.
Additionally, comparing different interaction models in terms of their proximity to the experimentally determined structure of proteins could help refine the modeling approach and guide future improvements in quantum \gls{psp}.
One interesting objective would be to integrate this method into a multi-scale modeling framework, where the coarse-grained structures predicted by the quantum workflow can serve as initial guesses for more detailed all-atom simulations, such as molecular dynamics~\cite{karplus2002molecular}.
This would allow a direct comparison of the predicted structures with experimentally determined ones, providing a more comprehensive assessment of the biological relevance of quantum-generated structures.

Finally, we note that the hybrid workflow proposed in this work would also benefit greatly from improved classical techniques.
For instance, the classical post-processing step introduced in Sec.~\ref{sec:energy_function} can be accelerated by leveraging efficient parallelization techniques and advanced \gls{hpc} infrastructures to scale the energy calculations to even large protein instances.
This naturally fits into the emerging paradigm of quantum-centric supercomputing~\cite{alexeev2024quantum, robledo2025chemistry}, potentially enabling the simulation and prediction of even larger and more complex proteins in the future.
Moreover, more sophisticated classical optimization solvers, such as those based on simulated annealing or branch-and-bound algorithms, including \texttt{Gurobi} and \texttt{COUENNE}, could enable a more efficient estimation of the ground state for benchmarking larger coarse-grained proteins, surpassing the limitations of the exhaustive search method employed in this study.

In conclusion, this work establishes a scalable framework for protein structure prediction using problem-agnostic, hardware-efficient ansatzes within hybrid quantum-classical workflows. 
By efficiently distributing computational tasks between quantum and classical resources, this method enables the simulation of larger protein sequences and the incorporation of higher-order interactions than previously possible.

\vfill\eject
\begin{acknowledgments}
We acknowledge support from the Knut and Alice Wallenberg Foundation through the Wallenberg Center for Quantum Technology (WACQT).
This research was supported in part by Wellcome Leap under the Quantum for Bio (Q4Bio) Program.
L.G.-{\'{A}.} further acknowledges support from the Swedish Foundation for Strategic Research (grant number FUS21-0063) and OpenSuperQ-Plus100 (101113946).

Thank you, Mårten Skogh and Pontus Vikstål, for your support during this work.

\end{acknowledgments}

\section*{Code availability}

The code used in this work is publicly available at:
\linebreak
\url{https://github.com/HannaLinn/Efficient-Quantum-PSP-with-HEA/tree/main}.

\appendix

\onecolumngrid
\section{Protein candidates}
\label{app:protein_candidates}

\begin{table}[ht]
\caption{\label{tab:protein_instances}\textbf{Protein candidates used in this study.} The protein instances studied, with their PDB Identification Code, amino acid sequence, number of amino acids, and number of qubits used for each lattice encoding: tetrahedral, \gls{bcc}, and \gls{fcc}, are listed. The 7-amino-acid Alpha-Bag Cell Peptide is abbreviated as ``ABCP'' to maintain consistency with other names.}
\begin{ruledtabular}
    \begin{tabular}{l l c ccc}
        \textbf{PDB-ID} & \multicolumn{1}{c}{\textbf{Sequence}} & \textbf{$N$} & \textbf{Tetrahedral} & \textbf{\gls{bcc}} & \textbf{\gls{fcc}} \\
        \hline
        \rule{0pt}{1em}4QXX & GNLVS & 5 & - & - & 10 \\
        2OL9 & SNQNNF & 6 & - & - & 14 \\
        ABCP & APRLRFY & 7 & - & 14 & - \\
        2M6C & GCVLYPWC & 8 & - & 17 & 22 \\
        1N9U & DRVYIHPFHL & 10 & - & 23 & - \\
        5AWL & YYDPETGTWY & 10 & 13 & 23 & 30 \\
        2K2R & DLDALLADLE & 10 & 13 & - & 30 \\
        2MZX & QYQFWKNFQT & 10 & 13 & - & 30 \\
        1IXU & FATMRYPSDSDE & 12 & 17 & - & 38 \\
        2N5R & VRRFDLLKRILK & 12 & 17 & 29 & 38 \\
        2L24 & IFGAIAGFIKNIW & 13 & 19 & 32 & 42 \\
        6Q08 & INWLKLGKKIIASL & 14 & 21 & - & - \\
        1K43 & RGKWTYNGITYEGR & 14 & 21 & 35 & 46 \\
        8T61 & RHYYKFNSTGRHYHYY & 16 & 25 & 41 & - \\
        8T63 & WHMWNTVPNAKQVIAA & 16 & 25 & - & - \\
        2NDC & GGLRSLGRKILRAWKKYG & 18 & 29 & - & - \\
        2NDE & IGLRGLGRKIALIHKKYG & 18 & 29 & - & - \\
        2JOF & DAYAQWLKDGGPSSGRPPPS & 20 & 33 & - & - \\
        8B1X & KKPGASLAALQALQALQAAQAAKKY & 25 & 43 & - & - \\
        6A8Y & YYHFWHRGVTKRSLSPHRPRHSRLQR & 26 & 45 & - & - \\
    \end{tabular}
\end{ruledtabular}
\end{table}

Protein candidates chosen for each lattice and the number of qubits needed to encode them are listed in Table~\ref{tab:protein_instances}. 
Each protein is identified by its PDB-ID, along with its amino acid sequence and length $N$.
The number of qubits depends on both the length of the amino acid sequence and the specific lattice encoding, as each lattice has a different coordination number, see Sec.~\ref{sec:lattice} in the main text.
Not all sequences are modeled on all three lattices due to the varying qubit requirements, which can exceed the capabilities of current quantum hardware and the exhaustive search method used to find the ground state for benchmarking.
A dash (–) indicates that a given sequence was not encoded using that particular lattice.
The selected set spans a wide range of sequence lengths and structural complexity, from small peptides of 5 amino acids to larger proteins of up to 26 amino acids, allowing us to evaluate the scalability and performance of the quantum workflow across diverse problem instances.

\section{Complete results of probability distributions}
\label{app:complete_results}

\begin{figure*}[t]
\centering
    \centering
    \includegraphics[width=0.8\linewidth]{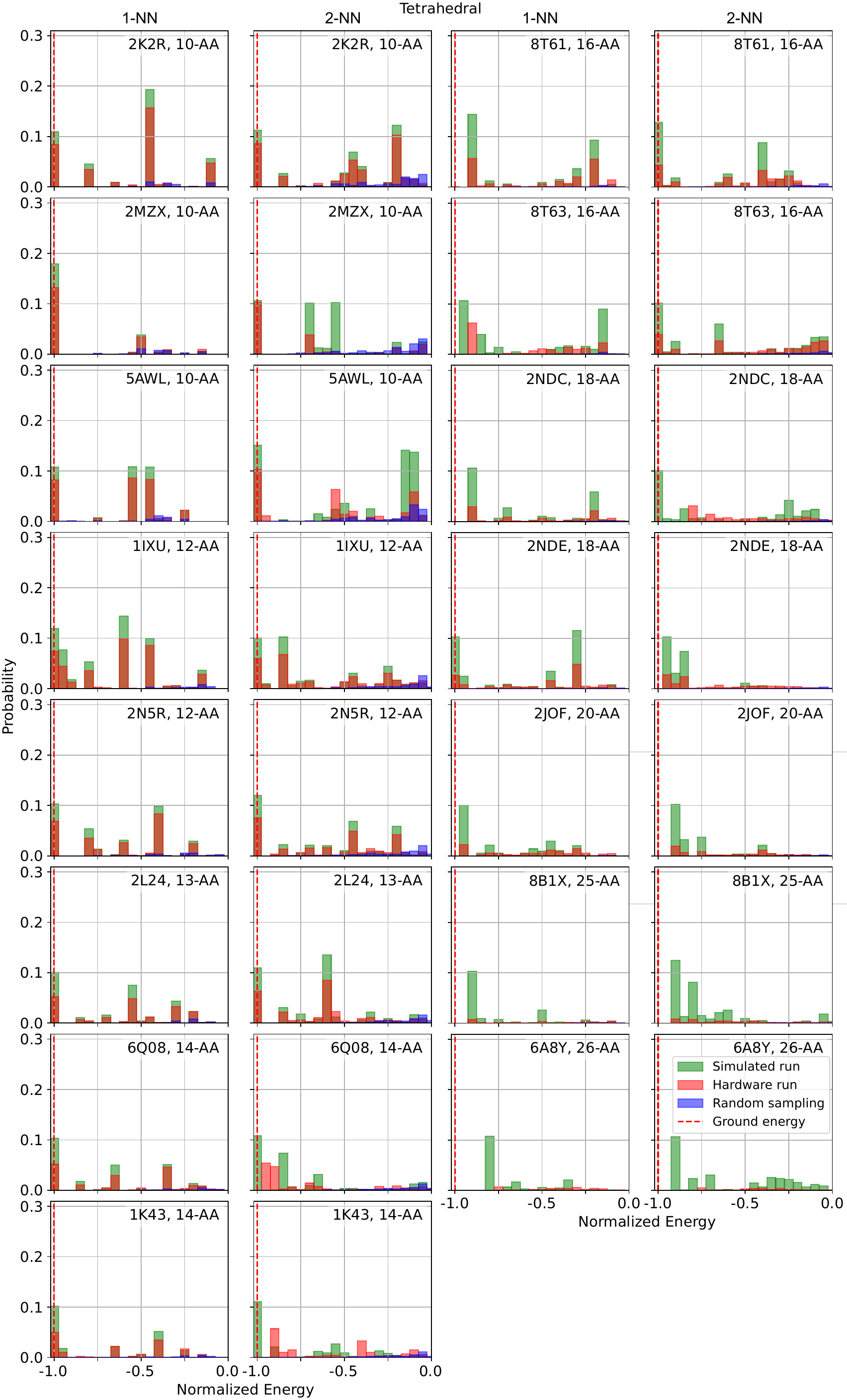}
\caption{\textbf{Complete energy distribution results for the tetrahedral lattice, including both 1-NN and 2-NN interactions.}}
\label{fig:add1}
\end{figure*}

\begin{figure}
    \centering
    \includegraphics[width=0.8\linewidth]{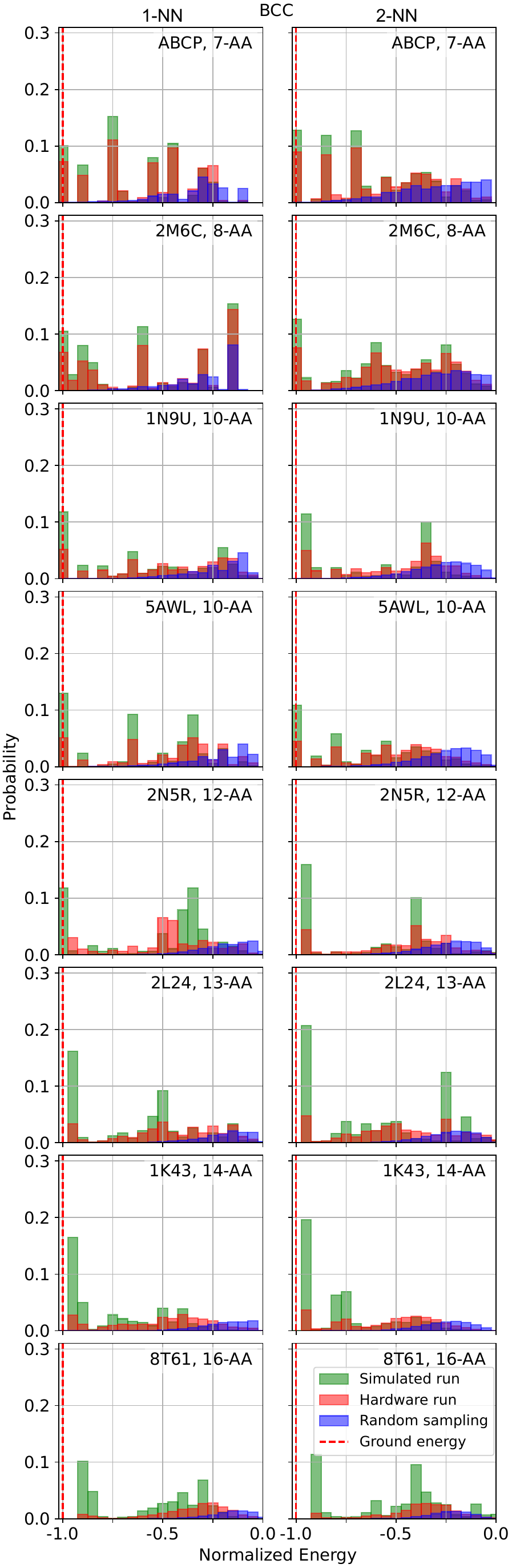}
    \caption{\textbf{Complete energy distribution results for the \gls{bcc} lattice, including both 1-NN (left) and 2-NN (right) interactions.}
    }
    \label{fig:add3}
\end{figure}

\begin{figure}[htbp]
    \centering
    \includegraphics[width=0.69\linewidth]{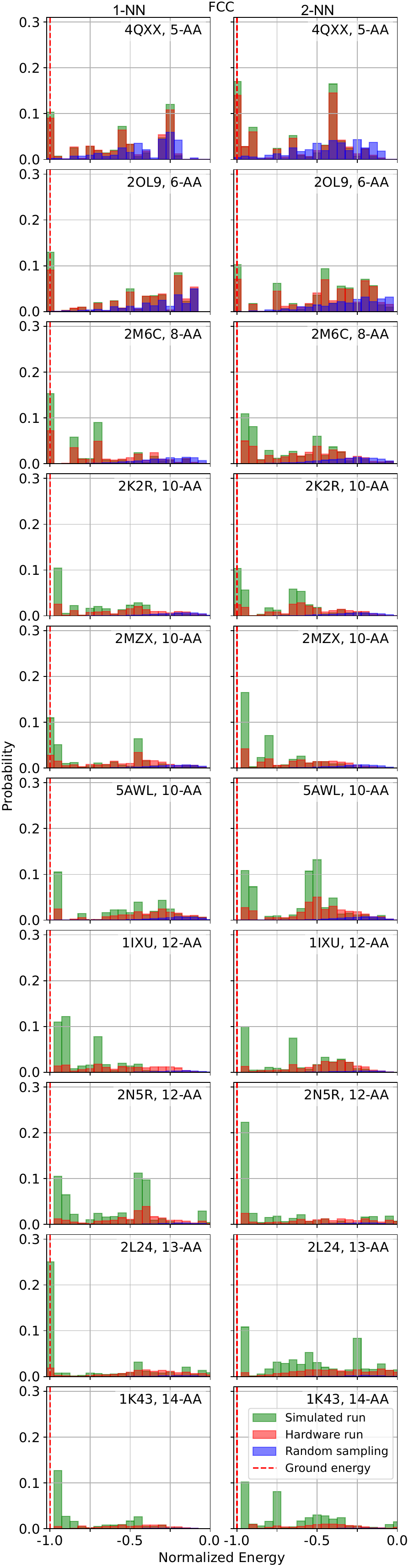}
    \caption{\textbf{Complete energy distribution results for the \gls{fcc} lattice, including both 1-NN (left) and 2-NN (right) interactions.}
    }
    \label{fig:add4}
\end{figure}

Complete results of the energy distributions of measured bitstrings from the optimized quantum circuits, selected as the lowest CVaR-cost instance among ten simulation runs and five hardware executions for each protein. 
These distributions highlight the likelihood of sampling states with energies near the ground state, across amino acid sequences identified by their PDB-IDs. 
Results are presented for three lattices---tetrahedral in Fig.~\ref{fig:add1}, \gls{bcc} in Fig.~\ref{fig:add3}, and \gls{fcc} in Fig.~\ref{fig:add4}---and for both 1-NN and 2-NN interaction models.

The x-axis represents the normalized energy, defined as the energy of a state divided by the absolute value of the ground state energy, with ground state energies indicated by red dashed lines. 
The y-axis represents the probability of energy windows sampled from the optimized circuit.
The noiseless simulation results are shown in green, while hardware results obtained from the \texttt{ibm\_kingston} device, featuring the Heron R2 quantum processor, are displayed as gray histogram bins.

\twocolumngrid

\clearpage
\bibliography{references}

\end{document}